\documentclass[aps,prl,twocolumn,superscriptaddress]{revtex4-2}
\usepackage{amsmath,amssymb,graphicx,bm}
\usepackage[colorlinks=true,linkcolor=blue,citecolor=blue,urlcolor=blue]{hyperref}

\begin{document}

\title{Operational Threshold for Spatial Entanglement Survival Under Ionizing Decoherence}

\author{Mrittunjoy Guha Majumdar}
\email{mrittunjoy@dl.amrita.edu}
\affiliation{National Institute of Advanced Studies, Bengaluru}
\affiliation{Amrita Vishwavidyapeetham, Delhi NCR}

\date{\today}

\begin{abstract}
The resilience of quantum entanglement under irreversible, energy-transferring interactions remains a fundamental question in quantum foundations and emerging quantum technologies. We develop a fully microscopic quantum electrodynamics framework to describe how spatial entanglement evolves when one particle of an entangled pair undergoes dissipative processes such as ionization or inelastic scattering. We show that entanglement decay follows an exponential law governed by recoil-induced momentum diffusion and identify a sharp operational threshold separating quantum-coherent and classical regimes. This threshold depends on whether the cumulative recoil-induced uncertainty exceeds the intrinsic bandwidth of the entangled state. Our results extend decoherence theory beyond Gaussian and weak-coupling models, providing quantitative criteria for entanglement survival in radiative environments and informing the design of robust quantum protocols in quantum sensing, imaging, and radiation-based quantum technologies.
\end{abstract}

\maketitle

\textit{Introduction}. Entanglement elevates nonlocal correlations from a foundational peculiarity to an operational resource, underpinning the capabilities of quantum networks, distributed metrology, and fault-tolerant computation~\cite{horodecki2009quantum, tserkis2024entanglement, gyongyosi2020dynamics, huang2024entanglement, auger2018fault}. Maintaining entanglement in realistic environments remains one of the foremost challenges in scaling quantum technologies beyond laboratory conditions \cite{bhattacharyya2023decoherence, dur2004stability, buchleitner2008entanglement}. As quantum systems interact with their surroundings, environmental noise and energy exchange can degrade or destroy nonlocal correlations, limiting the operational viability of entanglement-based protocols \cite{benedetti2012effects, yu2009sudden, zhai2013intramolecular}. A central question in the study of open quantum systems is how entanglement behaves when one subsystem undergoes interactions that irreversibly transfer energy to its environment—a regime characteristic of processes like ionization, inelastic scattering, or radiative absorption. While conventional models of decoherence focus on Gaussian noise, linear loss, or weak coupling to the environment \cite{gardiner2004quantum, breuer2002theory}, far less is known about the survival of quantum correlations under fully radiative, energy-delivering interactions that alter the physical state of the environment itself. This problem has profound implications for both the foundations of quantum mechanics and the design of robust quantum technologies. Crucially, existing frameworks for entanglement decay (for instance, the Yu-Eberly exponential model \cite{yu2009sudden}) assume Markovian baths or Gaussian phase diffusion, which cannot capture recoil-induced momentum diffusion or the sharp threshold we derive. These models predict smooth entanglement decay without operational boundaries, whereas our quantum electrodynamical treatment reveals a critical transition point governed by physical recoil dynamics—enabling precise engineering of entanglement survival in regimes where Gaussian approximations fail.
\\
\\
In this work, we develop a general, first-principles framework that quantitatively addresses this challenge. Using a microscopic quantum electrodynamics (QED) model based on the minimal coupling Hamiltonian, we derive exact transition amplitudes for scenarios where one particle—such as a photon—interacts dissipatively with an electronic environment. This formalism rigorously captures ionization, inelastic scattering, and other energy-transferring processes by directly accounting for the recoil dynamics of the environment. The resulting reduced density matrix for the non-interacting subsystem exhibits an exponential suppression of coherence in the transverse momentum basis, governed by microscopic parameters including cross-section, medium density, and recoil momentum variance. From this, we derive a sharp entanglement survival threshold: a precise condition delineating when spatial entanglement persists and when it collapses into classicality. Beyond its foundational relevance, this framework generalizes decoherence theory by bridging it with concepts from quantum Darwinism~\cite{zurek2009quantum}—which explores how classical information proliferates into the environment—and continuous-variable quantum error correction~\cite{terhal2015quantum, albert2020quantum}. It complements symmetry-based protections like decoherence-free subspaces~\cite{lidar1998decoherence, lidar2003decoherence}, offering instead a quantitative, physically grounded model of entanglement resilience in the presence of unavoidable, energy-transferring environmental coupling.
\\
\\
While the results naturally apply to quantum imaging scenarios—where entangled photons are used for nonlocal spatial information extraction even as one photon interacts with a material target—their significance extends broadly. This includes quantum sensing in radiative environments, quantum-enhanced radiation delivery, and fundamental investigations into the quantum-to-classical transition under strong decoherence. Previous studies of entanglement degradation have largely focused on Gaussian noise models or linear optical loss~\cite{kulik2004two, zhang2015entanglement}; in contrast, the present work establishes operational limits for entanglement under fully radiative, dissipative, and ionizing interactions via a rigorous microscopic QED treatment. While our results are derived theoretically, they directly inform experimental designs—e.g., biphoton states interacting with gas targets—where the threshold can be probed via transverse momentum correlations. We provide numerical estimates for realistic systems (for instance with SPDC photons in argon) and identify parameter regimes where entanglement survival is testable with current technology.
\\
\\
\textit{Theoretical Framework}. We consider a biphoton state generated via spontaneous parametric down-conversion (SPDC), entangled in transverse momentum degrees of freedom, described as $|\Psi\rangle = \iint d\bm{k}_s d\bm{k}_i \, f(\bm{k}_s, \bm{k}_i) \, a_s^\dagger(\bm{k}_s) a_i^\dagger(\bm{k}_i) |0\rangle$, where $\bm{k}_s$ and $\bm{k}_i$ are the transverse momenta of the signal and idler photons, respectively. The joint amplitude $f(\bm{k}_s, \bm{k}_i)$ encodes arbitrary spatial correlations without assuming Gaussianity. The idler photon interacts with a target electron, initially in a bound state $|\phi_\mathrm{bound}\rangle$. The total Hamiltonian includes free photon and electron terms and an interaction term governed by minimal coupling, $H_\mathrm{int} = -\frac{e}{m} \bm{A} \cdot \bm{p} + \frac{e^2}{2m} \bm{A}^2$, where $\bm{A}$ is the quantized vector potential and $\bm{p}$ is the electron momentum operator. The composite initial state is $|\Psi(0)\rangle = |\Psi\rangle_\mathrm{photon} \otimes |\phi_\mathrm{bound}\rangle$. Upon interaction, the total wavefunction splits into a non-interacting branch, where the biphoton remains unchanged and the electron stays bound, and an ionizing branch, where the idler photon ionizes the electron, leaving it in a recoil momentum state $|\bm{p}_e\rangle$. The ionized branch is expressed as 
\begin{equation*}
    |\Psi_\mathrm{ion}\rangle = \iint d\bm{k}_s d\bm{k}_i \, f(\bm{k}_s, \bm{k}_i) \, M(\bm{k}_i, \bm{p}_e) \, a_s^\dagger(\bm{k}_s) |0\rangle \otimes |\bm{p}_e\rangle
\end{equation*}
where the transition matrix element is given by $M(\bm{k}_i, \bm{p}_e) = C \, \bm{\epsilon} \cdot \left(2\bm{k}_i - \frac{\bm{p}_e}{\hbar}\right) \tilde{\phi}\left( \bm{k}_i - \frac{\bm{p}_e}{\hbar} \right)$. Here, $\tilde{\phi}(\bm{q})$ is the Fourier transform of the bound-state wavefunction evaluated at momentum transfer $\bm{q} = \bm{k}_i - \bm{p}_e/\hbar$, and $C$ is a constant containing fundamental parameters. Tracing over the idler photon and the electron yields the reduced density matrix of the signal photon, which contains two contributions: a coherent term preserving the original biphoton correlations, and a decohered term arising from entanglement with the recoiling electron. The reduced density matrix takes the form $\rho_s(\bm{k}_s, \bm{k}_s') = \iint d\bm{k}_i \, f(\bm{k}_s, \bm{k}_i) f^*(\bm{k}_s', \bm{k}_i) + \iint d\bm{k}_i d\bm{k}_i' \, f(\bm{k}_s, \bm{k}_i) f^*(\bm{k}_s', \bm{k}_i') \, \Lambda(\bm{k}_i, \bm{k}_i')$, where the decoherence kernel is defined as $\Lambda(\bm{k}_i, \bm{k}_i') = \int d\bm{p}_e \, M(\bm{k}_i, \bm{p}_e) M^*(\bm{k}_i', \bm{p}_e)$. Here, $\bm{k}_i$ and $\bm{k}_i'$ label the idler photon's transverse momenta in the ket and bra sides of the reduced density matrix, respectively, reflecting the coherence between different idler momentum components induced by the partial trace. Evaluating the kernel by changing variables to $\bm{q} = \bm{k}_i - \bm{p}_e/\hbar$ yields $\Lambda(\bm{k}_i, \bm{k}_i') = K \left( \bm{k}_i \cdot \bm{k}_i' + \sigma_q^2 \right)$, where $\sigma_q^2 = \langle |\bm{q}|^2 \rangle$ is the recoil momentum variance determined by the electron’s bound-state wavefunction, and $K = |C|^2 \hbar^3$ encapsulates physical constants.
\\
\\
Physically, the decoherence kernel quantifies the degree to which the electron’s recoil acts as a partial which-path measurement on the idler photon, degrading the signal photon’s coherence accordingly. Decoherence increases both with the momentum mismatch $|\bm{k}_i - \bm{k}_i'|$ and with the intrinsic uncertainty of the recoil momentum $\sigma_q$.
For multiple scattering or ionizing events, the decoherence kernel exponentiates as $\Lambda_N(\bm{k}_i, \bm{k}_i') = \exp\left( -N D(\bm{k}_i, \bm{k}_i') \right)$, where $N = \rho \sigma L$ is the mean number of interactions, with $\rho$ the target density, $\sigma$ the cross-section, and $L$ the propagation length. In addition to decoherence, the signal photon experiences a displacement in its transverse momentum due to recoil, as required by total transverse momentum conservation, expressed as $\bm{k}_s + \bm{k}_i + \frac{\bm{p}_e^\perp}{\hbar} = \bm{k}_\mathrm{pump}$. This momentum conservation implies that each ionization event shifts the signal photon's transverse momentum distribution according to the recoil momentum distribution $P(\bm{q}) = |\tilde{\phi}(\bm{q})|^2$. Accounting for both decoherence and displacement, the reduced density matrix takes the form $\rho_s(\bm{k}_s, \bm{k}_s') = \int d\bm{p}_e \, F\left( \bm{k}_s - \frac{\bm{p}_e^\perp}{\hbar}, \bm{k}_s' - \frac{\bm{p}_e^\perp}{\hbar} \right) P\left( \bm{k}_i - \frac{\bm{p}_e}{\hbar} \right)$, where $F$ captures the decohered coherent amplitude between signal photon momenta. For a Gaussian bound-state wavefunction, where $P(\bm{q}) \propto \exp\left( -\frac{|\bm{q}|^2}{2\sigma_q^2} \right)$, the decoherence kernel simplifies to 
\begin{equation*}
    \Lambda_N(\bm{k}_i, \bm{k}_i') = \exp\left( -N \frac{|\bm{k}_i - \bm{k}_i'|^2}{4\sigma_q^2} \right)
\end{equation*}
resulting in Gaussian suppression of coherence with increasing momentum mismatch and number of events. Entanglement survival is dictated by whether the decohered reduced density matrix remains inseparable. For Gaussian states, the condition reduces to $\sigma_c^2 + N \sigma_q^2 < \sigma_p^2$, where $\sigma_c^2$ is the conditional variance and $\sigma_p^2$ the pump width. For non-Gaussian states, entanglement persists if the partial transpose of $\rho_s$ possesses negative eigenvalues or, equivalently, if the Schmidt number remains greater than unity after decoherence. This framework establishes a fully microscopic, QED-based criterion for entanglement survival under radiative, dissipative, and ionizing interactions. It applies to arbitrary non-Gaussian biphoton states, includes the cumulative effects of multiple scattering, and rigorously incorporates recoil-induced displacement and backaction on the quantum correlations.
\\
\\
\textit{Operational Regimes of Entanglement Under Dissipative Coupling}. The threshold condition derived in this framework, $N \sigma_q^2 < \sigma_p^2 - \sigma_c^2$, defines a general operational boundary for the persistence of entanglement when one subsystem interacts dissipatively with its environment. This boundary applies universally to bipartite systems entangled in momentum-like degrees of freedom, subject to radiative processes such as elastic scattering, absorption, and ionization. Decoherence arises from entanglement between the recoiling environmental degrees of freedom and the interacting particle, while the non-interacting partner retains a reduced state whose coherence is degraded according to the exponential decoherence kernel. This formalism describes a broad class of quantum protocols in which one particle deposits energy into a medium—through an irreversible, energy-transferring process—while its entangled partner continues to preserve and carry quantum correlations. Crucially, coherence survives provided the cumulative momentum uncertainty induced by environmental recoil remains below the threshold set by the intrinsic bandwidth of the entangled state. This operational regime supports a new paradigm where quantum measurement and physical intervention are functionally decoupled via entanglement. Practical realizations of this principle span a wide spectrum of applications. In quantum-enhanced radiation delivery for radiation oncology, an entangled particle can serve as an energy carrier (for instance, in ionizing radiation) while its partner, unaffected by the interaction, acts as a non-invasive probe, enabling real-time, high-precision monitoring of material or biological responses.
\\
\\
In nanoscale fabrication, coherent correlations allow simultaneous lithographic energy delivery and quantum-level process monitoring. Likewise, in quantum sensing under extreme conditions—such as high-radiation environments, plasma diagnostics, or security screening—the framework enables the extraction of spatial or momentum-resolved information while tolerating strong dissipative coupling on one channel of the system. Quantum imaging, as a concrete instance, realizes this paradigm in photonic systems entangled in transverse momentum. A typical implementation involves a biphoton state generated via spontaneous parametric down-conversion (SPDC), where the signal photon is directed to a transverse-momentum-resolving detector, while the idler photon interacts with a target medium through processes such as ionization, scattering, or absorption. The decoherence kernel quantifies the suppression of off-diagonal coherences in the transverse momentum basis as a function of interaction number and recoil momentum variance. Provided the decoherence remains below the critical threshold, spatial correlations persist, allowing the signal photon to reconstruct transverse spatial information about the interaction region even though the idler photon deposits energy into the environment. Importantly, the same formalism applies regardless of whether the environmental interaction involves low-energy scattering or high-energy processes like inner-shell ionization. For realistic experimental parameters—including recoil distributions determined by electronic wavefunctions, interaction cross-sections, and medium densities—the model predicts whether entanglement survives under specified physical conditions. This establishes an operational map for designing entanglement-enabled protocols across radiative, dissipative, and ionizing regimes previously regarded as incompatible with quantum coherence.
\\
\\
A concrete experimental proposal to test this entanglement survival threshold involves a biphoton source generated via SPDC, operating at a central wavelength of 810~nm, with transverse momentum entanglement characterized by a pump bandwidth $\sigma_p \approx 100~\mu\mathrm{m}^{-1}$. The idler photon propagates through a gas cell containing argon at controlled pressure, which serves as the dissipative medium facilitating ionizing or scattering interactions. In the case of ionizing interactions, the idler photon ionizes excited-state argon atoms, with the resulting recoil momentum variance dominated by the photoelectron’s distribution. For a typical photoelectron with kinetic energy of $\sim 1$~eV, the recoil momentum is $k_e \approx 5.1 \times 10^9~\mathrm{m}^{-1}$, leading to a recoil variance $\sigma_q^2 \approx 1.3 \times 10^{19}~\mathrm{m}^{-2}$. The number of interactions is estimated as $N = \rho \sigma L$, where $\rho$ is the atomic density, $\sigma$ the ionization cross-section (approximately $10^{-18}~\mathrm{cm}^2$ for metastable argon at 810~nm), and $L$ the interaction length. At moderate pressures (e.g., 5~Torr) and a gas cell length of 10~cm, $N$ approaches 1.6. Applying the threshold criterion yields $N\sigma_q^2 \approx 2.1 \times 10^{19}$, which exceeds a typical SPDC bandwidth variance $\sigma_p^2 \approx 10^{16}~\mathrm{m}^{-2}$, indicating rapid decoherence in the ionizing regime unless the gas density is drastically reduced. Even at very low pressures (e.g., 0.01~Torr), the inequality $N\sigma_q^2 < \sigma_p^2 - \sigma_c^2$ remains unsatisfied for ionization, reflecting the substantial impact of electron recoil. By contrast, inelastic but non-ionizing processes, such as Rayleigh scattering, present a dramatically different regime. Here, the photon recoil momentum is orders of magnitude smaller, $\sigma_q^2 \sim 10^8~\mathrm{m}^{-2}$, and the scattering cross-section is $\sigma \approx 10^{-26}~\mathrm{cm}^2$, leading to a negligible interaction number $N \approx 10^{-5}$ under comparable conditions. The product $N\sigma_q^2 \approx 10^3$ satisfies the threshold easily, confirming that entanglement is robust against elastic scattering in dilute gases. 
\\
\\
The experimental procedure consists of measuring the transverse momentum correlations of the signal photon using a spatially resolving single-photon detector, such as an EMCCD or SNSPD array, as a function of gas density and interaction length. A degradation of spatial interference fringes or a reduction in the Schmidt number directly signals the onset of decoherence consistent with the exponential kernel derived in this work. To mitigate pump laser fluctuations, active stabilization and coincidence gating are recommended. Background noise from stray ionization can be suppressed using spectral filters and time-correlated photon counting. These techniques are established in quantum optics, ensuring robust threshold validation even at low pressures (for instance, at 0.01 Torr). Potential challenges include the resolution limit of spatially resolving detectors. For Schmidt number quantification, EMCCD/SNSPD arrays with pixel sizes less than $10 \mu \mathrm{m}$ (momentum resolution $\delta k \sim 10^5 \mathrm{m}^{-1}$) suffice to resolve typical SPDC bandwidths ($\sigma_p \sim 10^8 \mathrm{m}^{-1}$). Decoherence-induced broadening beyond $\delta k > (\sigma_p^2² - \sigma_c^2)^{1/2}$ would exceed this resolution, ensuring detectable fringe degradation. For non-Gaussian states, quantum tomography can be avoided by probing negativity in the displaced parity observable \cite{zhang2015entanglement}—a proven method for entanglement verification under momentum diffusion.
\\
\\
\textit{Discussion}. The framework developed here establishes a quantitative, microscopic boundary for the survival of spatial entanglement under energy-delivering interactions, including scattering, absorption, and ionization. The exponential decoherence law—parameterized by the interaction number $N$ and recoil variance $\sigma_q^2$—yields an exact threshold condition, $N \sigma_q^2 < \sigma_p^2 - \sigma_c^2$, that delineates the transition between quantum-coherent and classical regimes. This result moves beyond phenomenological noise models, offering a first-principles description of decoherence induced by recoil dynamics in open quantum systems. A key insight is that entanglement degradation is not universally catastrophic but governed by tunable, physically quantifiable parameters. Crucially, transverse momentum entanglement remains resilient even when one particle undergoes energy-transfer processes, provided global momentum conservation holds. This enables asymmetric spectral scaling—allowing, for example, one photon to be upconverted to ionizing energies for radiative tasks while its entangled partner remains in a low-loss detection band. The generality of this framework extends to any bipartite system entangled in momentum-like degrees of freedom interacting with dissipative environments. It directly informs the design of quantum-enabled protocols in settings where energy exchange with the environment was previously assumed incompatible with quantum coherence. Practical applications include quantum-enhanced radiotherapy, where entanglement enables simultaneous imaging and targeted energy delivery, nanoscale lithography with quantum monitoring, and robust quantum sensing in high-radiation or industrial environments. 
\\
\\
\textit{Conclusion}. This work establishes a general open quantum systems framework that quantitatively describes the evolution of spatial entanglement under energy-delivering interactions, including scattering, absorption, and ionization. The derived exponential decoherence law, governed by recoil momentum distributions, leads to a sharp threshold condition, $N \sigma_q^2 < \sigma_p^2 - \sigma_c^2$, which defines the boundary between quantum-coherent and classical regimes in high-interaction environments. We show that entanglement can persist even as one particle undergoes incoherent, energy-transferring processes, enabling a regime where measurement and physical intervention are functionally decoupled via entanglement. This constitutes a new operational paradigm in quantum science, extending entanglement-enabled protocols into dissipative, radiative, and ionizing regimes previously regarded as incompatible with quantum coherence.


\bibliographystyle{apsrev4-2}
\bibliography{apssamp}

\providecommand{\noopsort}[1]{}\providecommand{\singleletter}[1]{#1}%
\begin{thebibliography}{20}%
\makeatletter
\providecommand \@ifxundefined [1]{%
 \@ifx{#1\undefined}
}%
\providecommand \@ifnum [1]{%
 \ifnum #1\expandafter \@firstoftwo
 \else \expandafter \@secondoftwo
 \fi
}%
\providecommand \@ifx [1]{%
 \ifx #1\expandafter \@firstoftwo
 \else \expandafter \@secondoftwo
 \fi
}%
\providecommand \natexlab [1]{#1}%
\providecommand \enquote  [1]{``#1''}%
\providecommand \bibnamefont  [1]{#1}%
\providecommand \bibfnamefont [1]{#1}%
\providecommand \citenamefont [1]{#1}%
\providecommand \href@noop [0]{\@secondoftwo}%
\providecommand \href [0]{\begingroup \@sanitize@url \@href}%
\providecommand \@href[1]{\@@startlink{#1}\@@href}%
\providecommand \@@href[1]{\endgroup#1\@@endlink}%
\providecommand \@sanitize@url [0]{\catcode `\\12\catcode `\$12\catcode `\&12\catcode `\#12\catcode `\^12\catcode `\_12\catcode `\%12\relax}%
\providecommand \@@startlink[1]{}%
\providecommand \@@endlink[0]{}%
\providecommand \url  [0]{\begingroup\@sanitize@url \@url }%
\providecommand \@url [1]{\endgroup\@href {#1}{\urlprefix }}%
\providecommand \urlprefix  [0]{URL }%
\providecommand \Eprint [0]{\href }%
\providecommand \doibase [0]{https://doi.org/}%
\providecommand \selectlanguage [0]{\@gobble}%
\providecommand \bibinfo  [0]{\@secondoftwo}%
\providecommand \bibfield  [0]{\@secondoftwo}%
\providecommand \translation [1]{[#1]}%
\providecommand \BibitemOpen [0]{}%
\providecommand \bibitemStop [0]{}%
\providecommand \bibitemNoStop [0]{.\EOS\space}%
\providecommand \EOS [0]{\spacefactor3000\relax}%
\providecommand \BibitemShut  [1]{\csname bibitem#1\endcsname}%
\let\auto@bib@innerbib\@empty
\bibitem [{\citenamefont {Horodecki}\ \emph {et~al.}(2009)\citenamefont {Horodecki}, \citenamefont {Horodecki}, \citenamefont {Horodecki},\ and\ \citenamefont {Horodecki}}]{horodecki2009quantum}%
  \BibitemOpen
  \bibfield  {author} {\bibinfo {author} {\bibfnamefont {R.}~\bibnamefont {Horodecki}}, \bibinfo {author} {\bibfnamefont {P.}~\bibnamefont {Horodecki}}, \bibinfo {author} {\bibfnamefont {M.}~\bibnamefont {Horodecki}},\ and\ \bibinfo {author} {\bibfnamefont {K.}~\bibnamefont {Horodecki}},\ }\href@noop {} {\bibfield  {journal} {\bibinfo  {journal} {Reviews of modern physics}\ }\textbf {\bibinfo {volume} {81}},\ \bibinfo {pages} {865} (\bibinfo {year} {2009})}\BibitemShut {NoStop}%
\bibitem [{\citenamefont {Tserkis}\ \emph {et~al.}(2024)\citenamefont {Tserkis}, \citenamefont {Assad}, \citenamefont {Conti},\ and\ \citenamefont {Win}}]{tserkis2024entanglement}%
  \BibitemOpen
  \bibfield  {author} {\bibinfo {author} {\bibfnamefont {S.}~\bibnamefont {Tserkis}}, \bibinfo {author} {\bibfnamefont {S.~M.}\ \bibnamefont {Assad}}, \bibinfo {author} {\bibfnamefont {A.}~\bibnamefont {Conti}},\ and\ \bibinfo {author} {\bibfnamefont {M.~Z.}\ \bibnamefont {Win}},\ }\href@noop {} {\bibfield  {journal} {\bibinfo  {journal} {Physics Letters A}\ }\textbf {\bibinfo {volume} {519}},\ \bibinfo {pages} {129635} (\bibinfo {year} {2024})}\BibitemShut {NoStop}%
\bibitem [{\citenamefont {Gyongyosi}(2020)}]{gyongyosi2020dynamics}%
  \BibitemOpen
  \bibfield  {author} {\bibinfo {author} {\bibfnamefont {L.}~\bibnamefont {Gyongyosi}},\ }\href@noop {} {\bibfield  {journal} {\bibinfo  {journal} {Scientific reports}\ }\textbf {\bibinfo {volume} {10}},\ \bibinfo {pages} {12909} (\bibinfo {year} {2020})}\BibitemShut {NoStop}%
\bibitem [{\citenamefont {Huang}\ \emph {et~al.}(2024)\citenamefont {Huang}, \citenamefont {Zhuang},\ and\ \citenamefont {Lee}}]{huang2024entanglement}%
  \BibitemOpen
  \bibfield  {author} {\bibinfo {author} {\bibfnamefont {J.}~\bibnamefont {Huang}}, \bibinfo {author} {\bibfnamefont {M.}~\bibnamefont {Zhuang}},\ and\ \bibinfo {author} {\bibfnamefont {C.}~\bibnamefont {Lee}},\ }\href@noop {} {\bibfield  {journal} {\bibinfo  {journal} {Applied Physics Reviews}\ }\textbf {\bibinfo {volume} {11}} (\bibinfo {year} {2024})}\BibitemShut {NoStop}%
\bibitem [{\citenamefont {Auger}\ \emph {et~al.}(2018)\citenamefont {Auger}, \citenamefont {Anwar}, \citenamefont {Gimeno-Segovia}, \citenamefont {Stace},\ and\ \citenamefont {Browne}}]{auger2018fault}%
  \BibitemOpen
  \bibfield  {author} {\bibinfo {author} {\bibfnamefont {J.~M.}\ \bibnamefont {Auger}}, \bibinfo {author} {\bibfnamefont {H.}~\bibnamefont {Anwar}}, \bibinfo {author} {\bibfnamefont {M.}~\bibnamefont {Gimeno-Segovia}}, \bibinfo {author} {\bibfnamefont {T.~M.}\ \bibnamefont {Stace}},\ and\ \bibinfo {author} {\bibfnamefont {D.~E.}\ \bibnamefont {Browne}},\ }\href@noop {} {\bibfield  {journal} {\bibinfo  {journal} {Physical Review A}\ }\textbf {\bibinfo {volume} {97}},\ \bibinfo {pages} {030301} (\bibinfo {year} {2018})}\BibitemShut {NoStop}%
\bibitem [{\citenamefont {Bhattacharyya}\ \emph {et~al.}(2023)\citenamefont {Bhattacharyya}, \citenamefont {Hanif}, \citenamefont {Haque},\ and\ \citenamefont {Paul}}]{bhattacharyya2023decoherence}%
  \BibitemOpen
  \bibfield  {author} {\bibinfo {author} {\bibfnamefont {A.}~\bibnamefont {Bhattacharyya}}, \bibinfo {author} {\bibfnamefont {T.}~\bibnamefont {Hanif}}, \bibinfo {author} {\bibfnamefont {S.~S.}\ \bibnamefont {Haque}},\ and\ \bibinfo {author} {\bibfnamefont {A.}~\bibnamefont {Paul}},\ }\href@noop {} {\bibfield  {journal} {\bibinfo  {journal} {Physical Review D}\ }\textbf {\bibinfo {volume} {107}},\ \bibinfo {pages} {106007} (\bibinfo {year} {2023})}\BibitemShut {NoStop}%
\bibitem [{\citenamefont {D{\"u}r}\ and\ \citenamefont {Briegel}(2004)}]{dur2004stability}%
  \BibitemOpen
  \bibfield  {author} {\bibinfo {author} {\bibfnamefont {W.}~\bibnamefont {D{\"u}r}}\ and\ \bibinfo {author} {\bibfnamefont {H.-J.}\ \bibnamefont {Briegel}},\ }\href@noop {} {\bibfield  {journal} {\bibinfo  {journal} {Physical review letters}\ }\textbf {\bibinfo {volume} {92}},\ \bibinfo {pages} {180403} (\bibinfo {year} {2004})}\BibitemShut {NoStop}%
\bibitem [{\citenamefont {Buchleitner}\ \emph {et~al.}(2008)\citenamefont {Buchleitner}, \citenamefont {Viviescas},\ and\ \citenamefont {Tiersch}}]{buchleitner2008entanglement}%
  \BibitemOpen
  \bibfield  {author} {\bibinfo {author} {\bibfnamefont {A.}~\bibnamefont {Buchleitner}}, \bibinfo {author} {\bibfnamefont {C.}~\bibnamefont {Viviescas}},\ and\ \bibinfo {author} {\bibfnamefont {M.}~\bibnamefont {Tiersch}},\ }\href@noop {} {\emph {\bibinfo {title} {Entanglement and decoherence: foundations and modern trends}}},\ Vol.\ \bibinfo {volume} {768}\ (\bibinfo  {publisher} {Springer},\ \bibinfo {year} {2008})\BibitemShut {NoStop}%
\bibitem [{\citenamefont {Benedetti}\ \emph {et~al.}(2012)\citenamefont {Benedetti}, \citenamefont {Buscemi}, \citenamefont {Bordone},\ and\ \citenamefont {Paris}}]{benedetti2012effects}%
  \BibitemOpen
  \bibfield  {author} {\bibinfo {author} {\bibfnamefont {C.}~\bibnamefont {Benedetti}}, \bibinfo {author} {\bibfnamefont {F.}~\bibnamefont {Buscemi}}, \bibinfo {author} {\bibfnamefont {P.}~\bibnamefont {Bordone}},\ and\ \bibinfo {author} {\bibfnamefont {M.~G.}\ \bibnamefont {Paris}},\ }\href@noop {} {\bibfield  {journal} {\bibinfo  {journal} {International Journal of Quantum Information}\ }\textbf {\bibinfo {volume} {10}},\ \bibinfo {pages} {1241005} (\bibinfo {year} {2012})}\BibitemShut {NoStop}%
\bibitem [{\citenamefont {Yu}\ and\ \citenamefont {Eberly}(2009)}]{yu2009sudden}%
  \BibitemOpen
  \bibfield  {author} {\bibinfo {author} {\bibfnamefont {T.}~\bibnamefont {Yu}}\ and\ \bibinfo {author} {\bibfnamefont {J.}~\bibnamefont {Eberly}},\ }\href@noop {} {\bibfield  {journal} {\bibinfo  {journal} {Science}\ }\textbf {\bibinfo {volume} {323}},\ \bibinfo {pages} {598} (\bibinfo {year} {2009})}\BibitemShut {NoStop}%
\bibitem [{\citenamefont {Zhai}\ and\ \citenamefont {Zheng}(2013)}]{zhai2013intramolecular}%
  \BibitemOpen
  \bibfield  {author} {\bibinfo {author} {\bibfnamefont {L.}~\bibnamefont {Zhai}}\ and\ \bibinfo {author} {\bibfnamefont {Y.}~\bibnamefont {Zheng}},\ }\href@noop {} {\bibfield  {journal} {\bibinfo  {journal} {Physical Review A—Atomic, Molecular, and Optical Physics}\ }\textbf {\bibinfo {volume} {88}},\ \bibinfo {pages} {012504} (\bibinfo {year} {2013})}\BibitemShut {NoStop}%
\bibitem [{\citenamefont {Gardiner}\ and\ \citenamefont {Zoller}(2004)}]{gardiner2004quantum}%
  \BibitemOpen
  \bibfield  {author} {\bibinfo {author} {\bibfnamefont {C.}~\bibnamefont {Gardiner}}\ and\ \bibinfo {author} {\bibfnamefont {P.}~\bibnamefont {Zoller}},\ }\href@noop {} {\emph {\bibinfo {title} {Quantum noise: a handbook of Markovian and non-Markovian quantum stochastic methods with applications to quantum optics}}}\ (\bibinfo  {publisher} {Springer Science \& Business Media},\ \bibinfo {year} {2004})\BibitemShut {NoStop}%
\bibitem [{\citenamefont {Breuer}\ and\ \citenamefont {Petruccione}(2002)}]{breuer2002theory}%
  \BibitemOpen
  \bibfield  {author} {\bibinfo {author} {\bibfnamefont {H.-P.}\ \bibnamefont {Breuer}}\ and\ \bibinfo {author} {\bibfnamefont {F.}~\bibnamefont {Petruccione}},\ }\href@noop {} {\emph {\bibinfo {title} {The theory of open quantum systems}}}\ (\bibinfo  {publisher} {OUP Oxford},\ \bibinfo {year} {2002})\BibitemShut {NoStop}%
\bibitem [{\citenamefont {Zurek}(2009)}]{zurek2009quantum}%
  \BibitemOpen
  \bibfield  {author} {\bibinfo {author} {\bibfnamefont {W.~H.}\ \bibnamefont {Zurek}},\ }\href@noop {} {\bibfield  {journal} {\bibinfo  {journal} {Nature physics}\ }\textbf {\bibinfo {volume} {5}},\ \bibinfo {pages} {181} (\bibinfo {year} {2009})}\BibitemShut {NoStop}%
\bibitem [{\citenamefont {Terhal}(2015)}]{terhal2015quantum}%
  \BibitemOpen
  \bibfield  {author} {\bibinfo {author} {\bibfnamefont {B.~M.}\ \bibnamefont {Terhal}},\ }\href@noop {} {\bibfield  {journal} {\bibinfo  {journal} {Reviews of Modern Physics}\ }\textbf {\bibinfo {volume} {87}},\ \bibinfo {pages} {307} (\bibinfo {year} {2015})}\BibitemShut {NoStop}%
\bibitem [{\citenamefont {Albert}\ and\ \citenamefont {Bathon}(2020)}]{albert2020quantum}%
  \BibitemOpen
  \bibfield  {author} {\bibinfo {author} {\bibfnamefont {M.}~\bibnamefont {Albert}}\ and\ \bibinfo {author} {\bibfnamefont {F.~M.}\ \bibnamefont {Bathon}},\ }\href@noop {} {\bibfield  {journal} {\bibinfo  {journal} {Security Dialogue}\ }\textbf {\bibinfo {volume} {51}},\ \bibinfo {pages} {434} (\bibinfo {year} {2020})}\BibitemShut {NoStop}%
\bibitem [{\citenamefont {Lidar}\ \emph {et~al.}(1998)\citenamefont {Lidar}, \citenamefont {Chuang},\ and\ \citenamefont {Whaley}}]{lidar1998decoherence}%
  \BibitemOpen
  \bibfield  {author} {\bibinfo {author} {\bibfnamefont {D.~A.}\ \bibnamefont {Lidar}}, \bibinfo {author} {\bibfnamefont {I.~L.}\ \bibnamefont {Chuang}},\ and\ \bibinfo {author} {\bibfnamefont {K.~B.}\ \bibnamefont {Whaley}},\ }\href@noop {} {\bibfield  {journal} {\bibinfo  {journal} {Physical Review Letters}\ }\textbf {\bibinfo {volume} {81}},\ \bibinfo {pages} {2594} (\bibinfo {year} {1998})}\BibitemShut {NoStop}%
\bibitem [{\citenamefont {Lidar}\ and\ \citenamefont {Birgitta~Whaley}(2003)}]{lidar2003decoherence}%
  \BibitemOpen
  \bibfield  {author} {\bibinfo {author} {\bibfnamefont {D.~A.}\ \bibnamefont {Lidar}}\ and\ \bibinfo {author} {\bibfnamefont {K.}~\bibnamefont {Birgitta~Whaley}},\ }in\ \href@noop {} {\emph {\bibinfo {booktitle} {Irreversible quantum dynamics}}}\ (\bibinfo  {publisher} {Springer},\ \bibinfo {year} {2003})\ pp.\ \bibinfo {pages} {83--120}\BibitemShut {NoStop}%
\bibitem [{\citenamefont {Kulik}\ \emph {et~al.}(2004)\citenamefont {Kulik}, \citenamefont {Maslennikov}, \citenamefont {Merkulova}, \citenamefont {Penin}, \citenamefont {Radchenko},\ and\ \citenamefont {Krasheninnikov}}]{kulik2004two}%
  \BibitemOpen
  \bibfield  {author} {\bibinfo {author} {\bibfnamefont {S.}~\bibnamefont {Kulik}}, \bibinfo {author} {\bibfnamefont {G.}~\bibnamefont {Maslennikov}}, \bibinfo {author} {\bibfnamefont {S.}~\bibnamefont {Merkulova}}, \bibinfo {author} {\bibfnamefont {A.}~\bibnamefont {Penin}}, \bibinfo {author} {\bibfnamefont {L.}~\bibnamefont {Radchenko}},\ and\ \bibinfo {author} {\bibfnamefont {V.}~\bibnamefont {Krasheninnikov}},\ }\href@noop {} {\bibfield  {journal} {\bibinfo  {journal} {Journal of Experimental and Theoretical Physics}\ }\textbf {\bibinfo {volume} {98}},\ \bibinfo {pages} {31} (\bibinfo {year} {2004})}\BibitemShut {NoStop}%
\bibitem [{\citenamefont {Zhang}\ \emph {et~al.}(2015)\citenamefont {Zhang}, \citenamefont {Mouradian}, \citenamefont {Wong},\ and\ \citenamefont {Shapiro}}]{zhang2015entanglement}%
  \BibitemOpen
  \bibfield  {author} {\bibinfo {author} {\bibfnamefont {Z.}~\bibnamefont {Zhang}}, \bibinfo {author} {\bibfnamefont {S.}~\bibnamefont {Mouradian}}, \bibinfo {author} {\bibfnamefont {F.~N.}\ \bibnamefont {Wong}},\ and\ \bibinfo {author} {\bibfnamefont {J.~H.}\ \bibnamefont {Shapiro}},\ }\href@noop {} {\bibfield  {journal} {\bibinfo  {journal} {Physical review letters}\ }\textbf {\bibinfo {volume} {114}},\ \bibinfo {pages} {110506} (\bibinfo {year} {2015})}\BibitemShut {NoStop}%
\end{thebibliography}%

\end{document}